\documentclass[aps,pre,onecolumn,12pt]{revtex4}
\usepackage{amsmath}
\usepackage{amssymb}
\usepackage{graphicx}
\usepackage{color}

\begin{document}
\title{Modular networks of word correlations on Twitter}
\author{Joachim Mathiesen, Pernille Yde and Mogens H. Jensen}
\affiliation{Niels Bohr Institute, University of Copenhagen, Blegdamsvej 17, DK-2100 Copenhagen, Denmark}

\begin{abstract}
Complex networks are important tools for analyzing the information flow in many aspects of nature and human society. Using data from the microblogging service Twitter, we study networks of correlations in the appearance of words from three different categories, international brands, nouns and US major cities. We create networks where the strength of links is determined by a similarity measure based on the rate of coappearance of words. In comparison with the null model, where words are assumed to be uncorrelated, the heavy-tailed distribution of pair correlations is shown to be a consequence of modules of words representing similar entities.
\end{abstract}
\maketitle
\section*{Introduction}
Networks are elegant representations of interactions between individuals in large communities and organizations\cite{B02, Borgatti2010, Kitsak2010}. These networks are constantly changing according to demands, fashions and flow of ideas\cite{RFFMV2010, Twitter2011, Huberman2009}. The worldwide popularity of social media such as Twitter\cite{KLPM10, Twitter2011, Huberman2009} have made them a considerable component in research on social networks\cite{king2010, C10}. Twitter is a microblogging service that allows registered users to post short text-based announcements, limited to 140 characters in length, known as ``tweets'', to an online stream. The frequency by which users interact on a global scale on Twitter allows for a high-resolution real-time analysis of movements in society.

From automatic queries to Twitter, we have estimated tweet rates of words from a given set $M$ containing selected words from one of the three different categories, international brand names, nouns and US major city names. The rate is measured by the number of new tweets posted per hour. For each query submitted at time $t$ about a specific word $a\in M$, Twitter returns a finite set of $n_a(t)$ tweets $\{T_1,\ldots,T_{n_a(t)}\}$. In additions to the text string $s$ containing the word $a$, each tweet contains the username of the author of the tweet, the time $t_i$ the tweet was posted and further details that we have not used. A tweet $T_i$ is therefore list of information $T_i=(s,t_i,\ldots)$. The maximum number of tweets returned from each query is $n_a=1500$. 

The time signal of tweets mentioning a specific word $a$, $\eta_a(t)$, can be written on the form
\begin{equation}
\eta(t)=\sum_i \delta(t-t_i),
\end{equation}
From the number of tweets and the timestamps we compute an averaged tweet rate of a word $a$,
\begin{equation}
\gamma_a(t)=\frac 1 {\tau} \int_{t_1}^{t_1+\tau}\eta(t) \mathrm{d} t  =\frac{n_a(t)}{\tau},  \label{tweetrate}
\end{equation}
Similarly we define a rate for two words $a$ and $b$ appearing in a tweet at the same time,
$\gamma_{ab}(t)={n_{ab}(t)}/{\tau}$.

Tweets containing words from the aforementioned categories were recorded over a period of 4 months November 2010 -- February 2011 and a period of two months January 2012 -- February 2012. In general the rate, at which new tweets appear containing words from each of the categories, is too high to count the total number of tweets. Our analysis is therefore based on estimated tweet rates computed from Eq. (\ref{tweetrate}) using $n_a=100$ and $n_a=1500$. When averaging over many queries, we did not see a significant difference our results when using different values of $n_a$.


We analyse the correlation between individual words within the said categories. For that purpose, we define a measure of similarity in terms of rate of co-appearance of pairs of words. The measure is then used to construct networks where links represent the degree of similarity. All three categories of words are shown to have a pronounced modularity. The way that we consider correlation networks can be seen as an alternative to existing studies on semantic networks (see e.g. \citep{ST05}).

\section*{Results}
We define a similarity measure between two words $a$ and $b$ in terms of the hourly average rate $\gamma_{ab}$ by which new tweets appear containing both $a$ and $b$. For example, by considering queries to Twitter containing the terms "Google" and "Microsoft", we get $\gamma_{Google}\approx 130000 $ tweets per hour and $\gamma_{Microsoft}\approx 17000$ tweets per hour whereas $\gamma_{Google,Microsoft}\approx 700$ tweets per hour (January 2011).
A normalized symmetric measure of similarity is naturally defined by
\begin{equation}
\omega_{ab}=\frac{\gamma_{a \cap b}}{\gamma_{a \cup b}}=\frac{\gamma_{ab}}{\gamma_{a}+\gamma_{b}-\gamma_{ab}}\label{measure}
\end{equation}
Alternatively one could use information theory to compute the similarity from the joint probability of observing two words in the same tweet\cite{CV07}. {This approach would have been useful had we had access to the normalized probabilities of observing $A$ and $B$. Here, because of limitations to the permissible sample rate of data we only have access to a fraction of the total number of posted tweets containing the relevant words and therefore we can only estimate the relative probabilities.}

In Fig.~\ref{fig2} we present networks of international brand names where the link strength is given by the measure Eq.~(\ref{measure}). {A threshold is introduced on the link strength in order to visualize primary structures, {\it i.e.}~links between brand with a similarity $\omega_{AB}<0.004$ are omitted.} We observe that the network is strongly modular with groups of brands representing similar products or services. As an example one can observe distinct groups of European car brands, Asiatic car brands, consulting and IT companies, and fashion brands. The modules in the network are coloured according to the community detection algorithm introduced in\cite{rosvall}. While most of the connections inside the modules are somewhat less surprising, the few links connecting the modules represent some less obvious relations between brands.

In Fig. \ref{cities}A, a similarity network of US cities is shown. The network provides an alternative map where individual cities only to some extent are grouped according to their geographical location. The network is dominated by a central module consisting of New York, Chicago, Atlanta, Los Angeles and Boston. This is not surprising as these cities are hubs in the American society. We observe a module of Californian cities that connects naturally to cities like Denver and Seattle. We also detect a module of east-coast to mid-western cities connecting to a module of southern cities. Again the modules were detected by the algorithm presented in \cite{rosvall}. It is natural to ask how much of the similarity between cities is influenced by the geographical distance between them. To answer this question, we have compared tweet rates with the distance between cities as well as the size of the cities. It turns out that there is a weak to moderate correlation between the size of a city and the number of tweets referring to that city. The co-appearance of two cities, however, has no clear correlation with their sizes and the distance between them. That said, when the nodes in the similarity network are arranged according to their geographical location it is evident that cities in same regions (states or neighbouring states) are better inter-connected and therefore often belong in the same module, see Fig. \ref{cities}B.

As a final example of a similarity network, we present in Fig.~\ref{fig2n} a network of nouns. From a list of 2000 common nouns in the English language, 200 nouns are randomly selected and the corresponding pairwise similarities are computed. Like the previous networks for brands and cities, the network of nouns also exhibits a pronounced modularity with modules e.g. representing similar food products.

{We now consider further the data underlying the link strength in the networks. As a main result, we obtain scale free distributions, 
\begin{equation}
p(\gamma_{ab})\sim\gamma_{ab}^{-\alpha},
\end{equation}
of the pairwise tweet rates $\gamma_{ab}$ over six orders of magnitude using the brand names, nouns as well as major cities, see Fig.~\ref{fig3}.} Surprisingly, the distributions all have the same scaling exponent $\alpha=1.40 \pm 0.02$ (s.d.). The distribution of the tweet rates of individual search terms $a$, $\gamma_a$, does not follow a clear scale invariant distribution (see inset of Fig.~\ref{fig3}). Moreover, the tweet rate of pairs $\gamma_{ab}$ does not follow trivially from the rate of the individual brands, that is, the rate is not proportional to the product $\gamma_a \gamma_b$ which would be the case if $a$ and $b$ were uncorrelated. In particular we notice that if the distribution of the rates $\gamma_x$ could be approximated by a scale invariant distribution $p(\gamma_x)\sim \gamma_x^{-\alpha}$ then the product $z=\gamma_a \gamma_b$ would follow a distribution
\begin{equation}
p(z)\sim z^{-\alpha}\log(z^2).
\end{equation}
which follows from introducing the auxiliary variable $v=\gamma_a/\gamma_b$ and performing the integral
\begin{equation}
\int_{\epsilon^2/z}^{z/\epsilon^2} p(z,v) \mathrm{d} v =\int_{\epsilon^2/z}^{z/\epsilon^2} p(\gamma_a(z,v),\gamma_b(z,v))\left|\frac{\partial (\gamma_a,\gamma_b)}{\partial(z,v)}\right| \mathrm{d} v,
\end{equation}
where $\epsilon$ is a characteristic minimum tweet-rate that we observe.

The logarithmic correction to the scaling does not provide a statistically significant fit to the data presented in Fig.~\ref{fig3}, that is a best fit has an exponent $\alpha\approx 2$ significantly larger than the tweet rate $\gamma_x$ of individual search terms (see the inset of Fig.~\ref{fig3}). A power-law distribution has also been observed for the co-occurrence of tags in social annotation systems \cite{CB09} where users annotate online resources such as web pages by lists of words. The exponent of the distribution in the annotation systems ($\alpha>2$) is larger than the one reported here and is close to the distribution of co-occurrence of nouns in sentences of novels considered below.

\begin{figure}[thb]
\centering
\includegraphics[width=.48\textwidth]{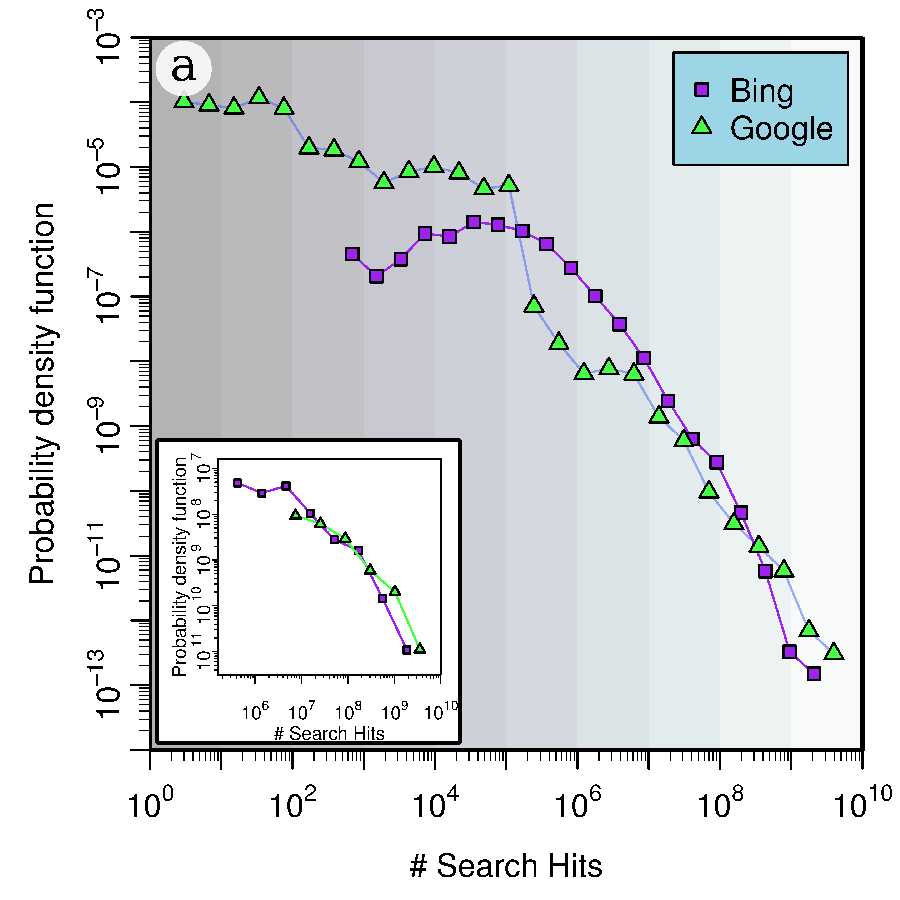}%
\includegraphics[width=.48\textwidth]{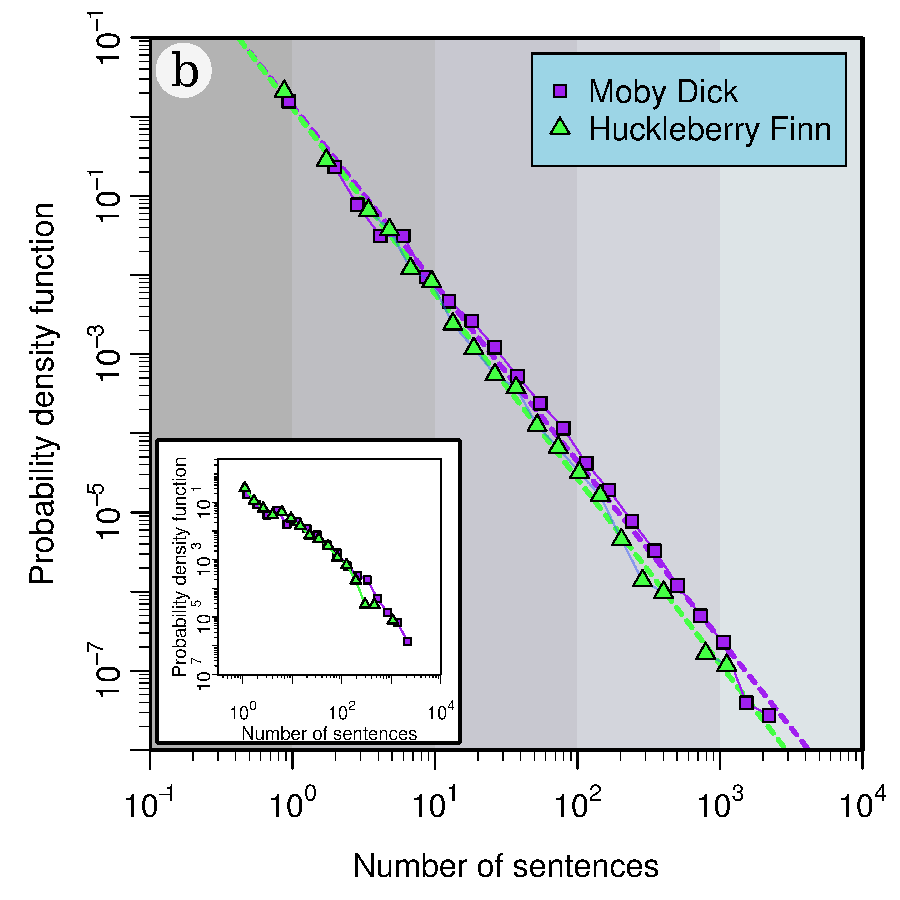}
\caption{{Probability density functions of the number of search hits returned from Bing and Google and for the number of sentences in which two nouns co-appear in novels.} In panel a) we performed pairwise queries on international brands to Bing and Google. In contrast to the result obtained from Twitter, we do not observe clear scale-free distributions. Inset: Probability density functions of search hits returned from queries on individual brands alone. Panel b) shows the number of sentences in which two nouns co-appear in the novels Huckleberry Finn (Mark Twain) and Moby-Dick (Herman Melville). The distributions are plotted on double-logarithmic scales and include the distributions of individual nouns. Dashed lines are best fit to a scale-free distribution and have exponents $\alpha=2.34\pm 0.04$(s.d.) (Huckleberry Finn) and $\alpha=2.24\pm 0.04$(s.d.) (Moby-Dick). Inset: Probability density function of the frequencies by which individual nouns appear in the same sentences.
}\label{fig4}
\end{figure}

\section*{Discussion}

For comparison, we have performed a similar analysis using search engines such as Google and Bing. The similarity between two words was computed from Eq.~(\ref{measure}) by inserting the number of web pages that the search engines return containing the words. That is, instead of a rate we now use a fixed number. {The distributions turn out to be significantly different (see Fig.~\ref{fig4}a) and do not show a clear scaling behavior as in the case of Twitter. This may in part be explained by the fact that the search engines return results from web pages which are not restricted in size and they cover a wide range of media.}

Finally, we compare the scaling behavior of word correlations observed on Twitter by considering the corresponding distribution of nouns in sentences of novels by Mark Twain (Huckleberry Finn) and Herman Melville (Moby-Dick). The sentences in these novels turn out to have a typical length comparable to the 140 character limit of a tweet and do indeed lead to broad but significantly steeper distributions in the word correlations (see Fig.~\ref{fig4}b). The novels are written by single authors and typically exhibits a more formal structure compared to the text messages. At the same time, the pair distribution of nouns are for the novels compatible with the null model where all words in the novels are randomized meaning that the correlated structures in the novels are rather weak. The distributions of individual words were considered for the same novels in \cite{BRM10}. Compared to the novels the distribution of the co-appearance of words in tweets is less broad, which might be because the active vocabulary of the average user of Twitter is less diverse than that of the authors of the two novels.

Scale invariance is often described by Zipf's law \citep{Z49} which states that the
frequency of a word (for instance in a language) is inversely proportional
to the rank in the frequency table. In its general formulation Zipf's law
says that the frequency $\gamma$ of a word is a power law in the rank
$\gamma \sim r^{-\alpha}$. For the corresponding probability density
functions we have 
$p(\gamma) d\gamma = p(r) dr \rightarrow p(\gamma) = p(r) \left|\frac{dr}{d\gamma}\right|$. Since $\frac{dk}{d\gamma} = \gamma^{-\frac{1+\alpha}{\alpha}}$ making the
natural assumption that the PDF of the rank is a constant, we obtain
the PDF of the frequency as
\begin{equation}
p(\gamma) \sim \gamma^{-\frac{1+\alpha}{\alpha}}
\end{equation}
Empirically the value $\alpha \sim 1$ has been found for words in a corpus of a natural language where as for the population size of cities $\alpha \sim 1.1$. In Fig.~\ref{fig4} (inset)
we observed a frequency distribution $p(\gamma) \sim \gamma^{-2}$ for words in two novels
leading to $\alpha \sim 1$ in good agreement with the 'established' Zipf result.
For Twitter sentences on the other hand we found $p(\gamma) \sim \gamma^{-1.4}$
leading to a rank exponent of the order $\alpha=2.5$ which is quite far
from the usual Zipf exponent. We thus conclude, that texts from human communication
on social media leads to a self-organized state that appears to have no resemblance with the structure of written texts.

Social media have become vital channels for advertising, the dissemination of news and spreading of political opinions, therefore an understanding of user communication in online social media provides important input not only to several branches of science but also for commercial purposes. For example, the value of a brand is determined by the consumer awareness and its apparent uniqueness. Companies put enormous efforts into positioning, i.e.~to create the right image in the mind of potential customers. The similarity measure and the corresponding correlation networks that we have suggested here, could be a first step in assessing market positions products. Finally, the universal correlations might provide basic information on the user awareness of e.g. brands and cities.


\section*{Acknowledgments}
Suggestions and comments by Alex Hunziker and Pengfei Tian are gratefully acknowledged. This study was supported by the Danish National Research Foundation through the Center for Models of Life.

\section*{Author Contributions}
J.M., P.Y and M.H.J. designed the research, performed the research, analyzed the data, and wrote the
paper

\section*{Competing financial interests}
The authors declare no competing financial interests.

\begin{figure}[!h]
\centering
\includegraphics[width=.45\textwidth]{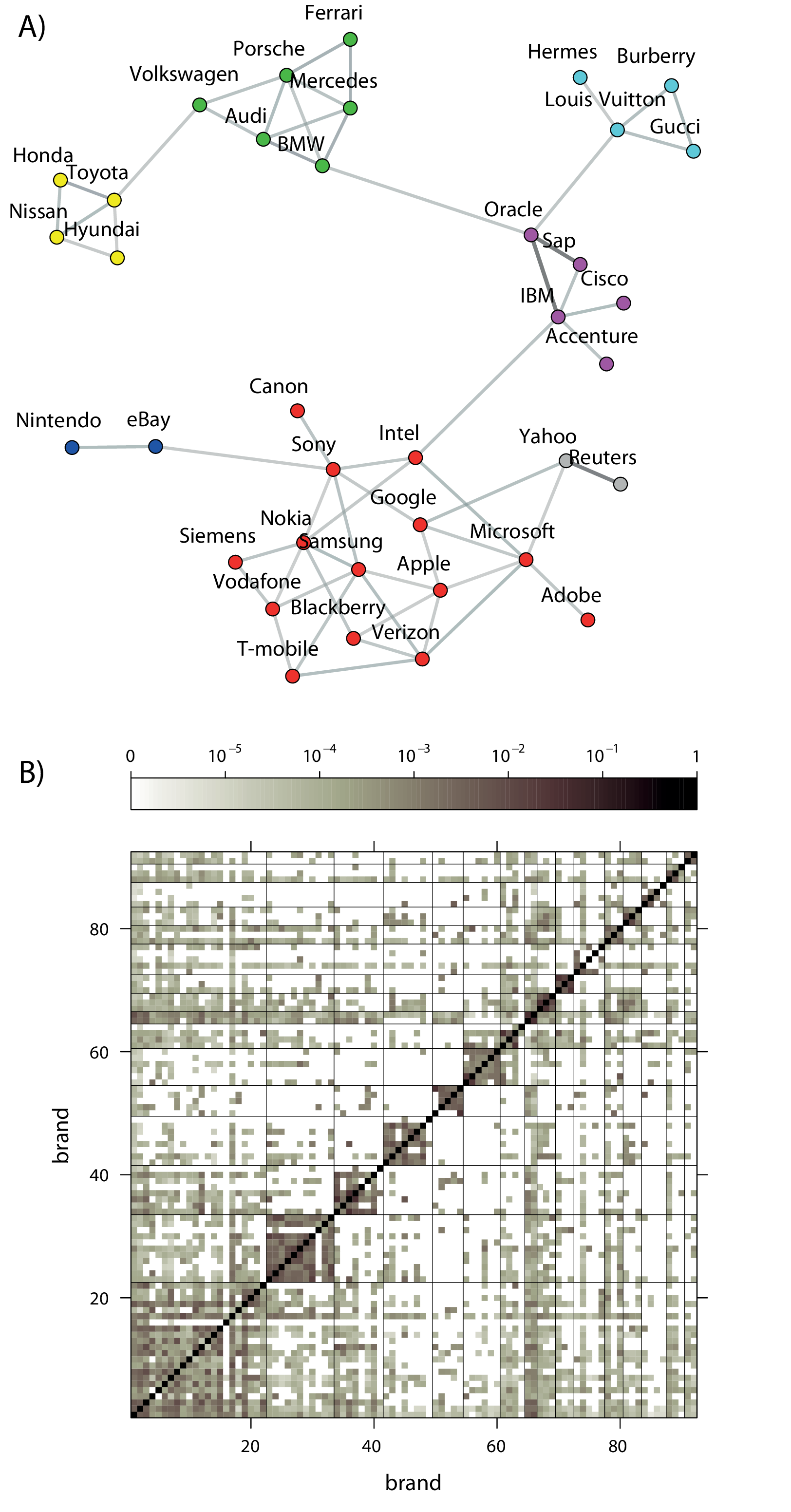}
\caption{{Network of correlations between international brands computed from the corresponding tweet rates on Twitter}. A link in the network represents the similarity measure computed using Eq. (\ref{measure}). Only links with a strength larger than 0.004 are shown. The color of the nodes represents modules of more inter-connected brands. Darker link colors mean stronger links.}\label{fig2}
\end{figure}

\begin{figure}[thb]
\centering
\includegraphics[width=.45\textwidth]{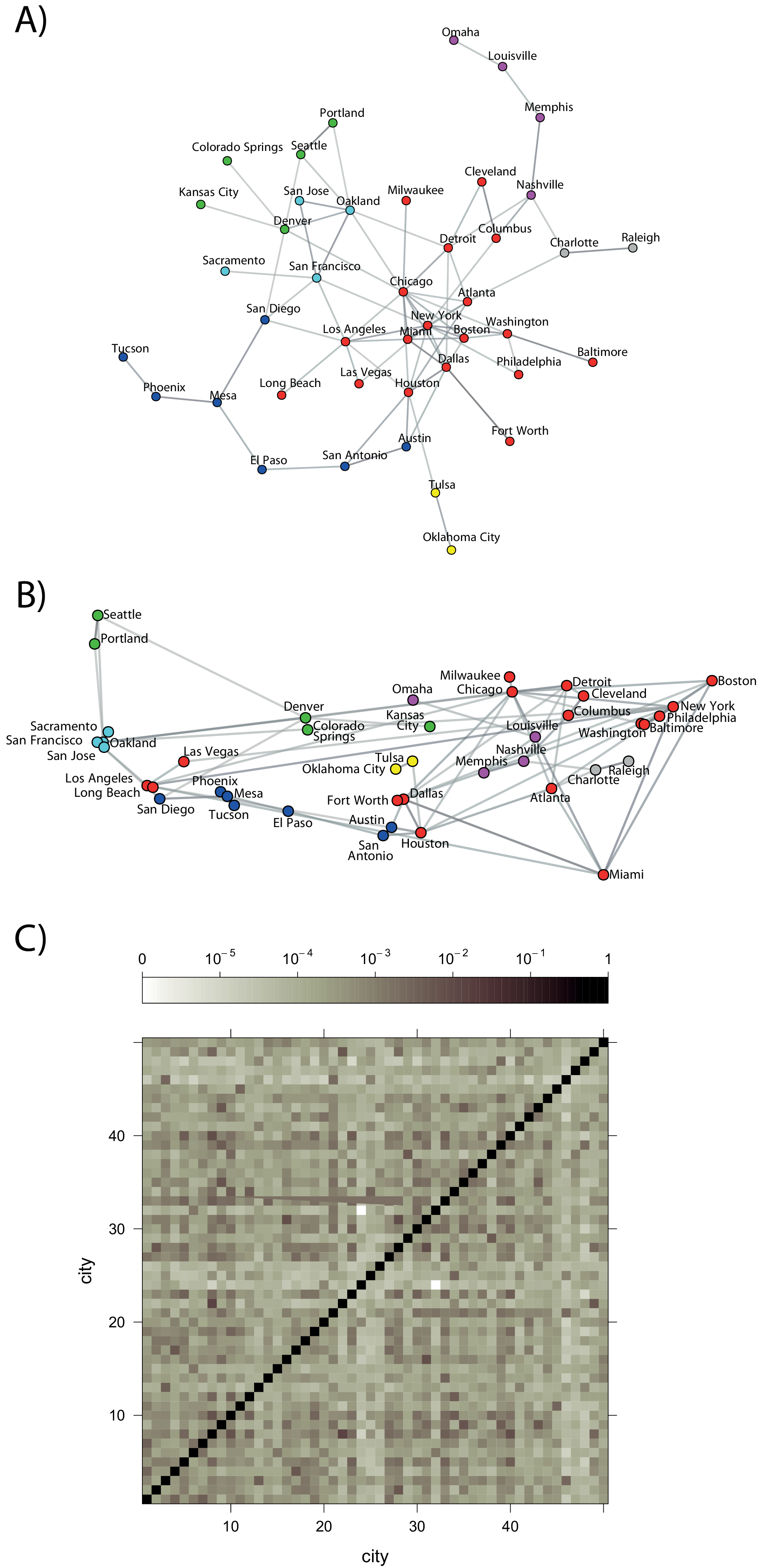}
\caption{{Network of cities with high similarity.} In panel A), we show a similarity network where nodes are located according to the algorithm of Fruchterman-Rheingold. In panel B), the corresponding network is shown where nodes are arranged according to the geographical location of the cities. In both panels only links with a strength larger than 0.004 are shown. In the network, darker link colors mean stronger links.}\label{cities}
\end{figure}

\begin{figure}[thb]
\centering
\includegraphics[width=.48\textwidth]{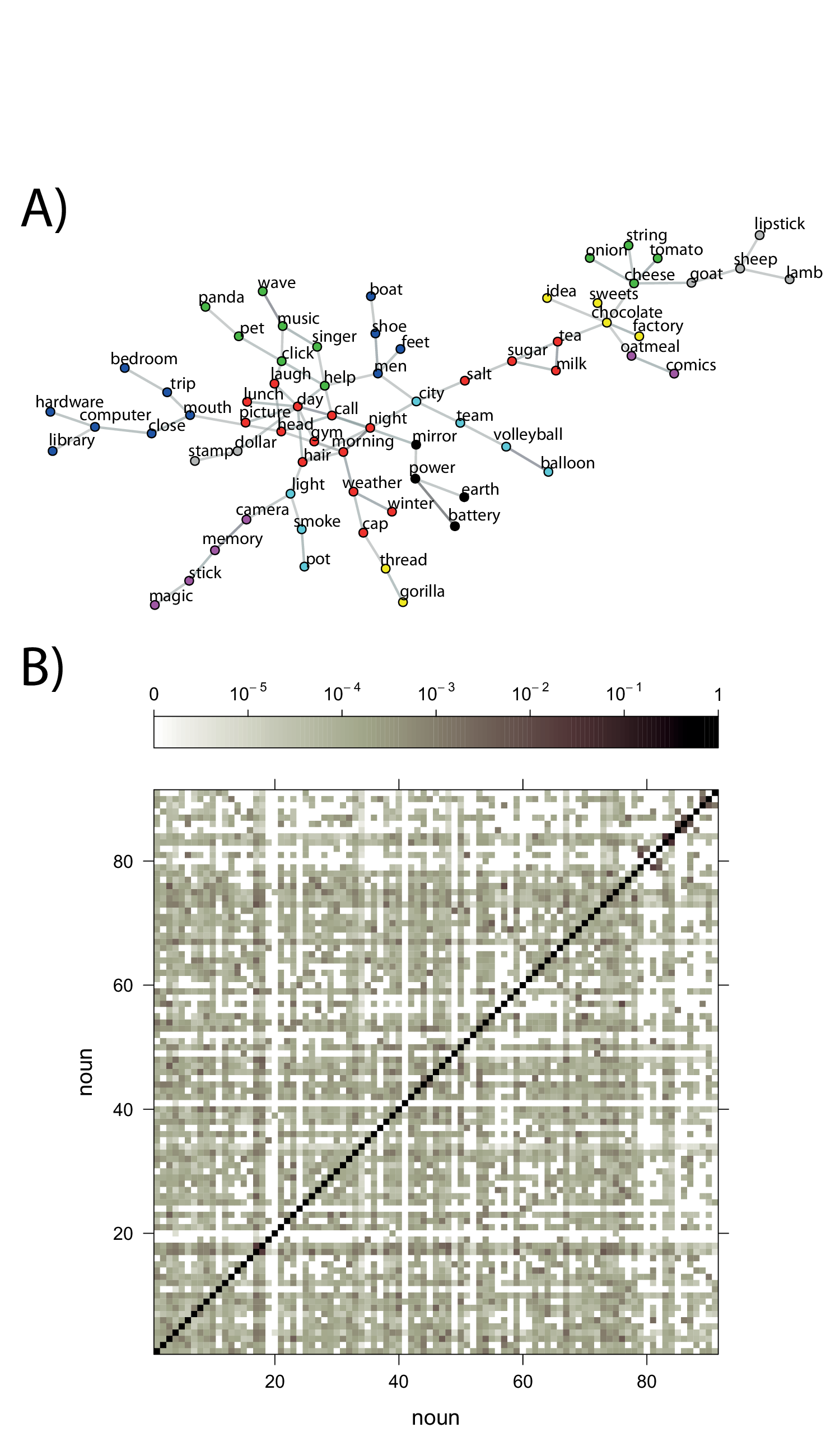}
\caption{{Network of nouns with high similarity.} Similarity network of 200 random nouns chosen from a list of the 2000 most common nouns. We only show the largest connected component for links with a strength larger than 0.04.}\label{fig2n}
\end{figure}

\begin{figure}[thb]
\centering
\includegraphics[width=.48\textwidth]{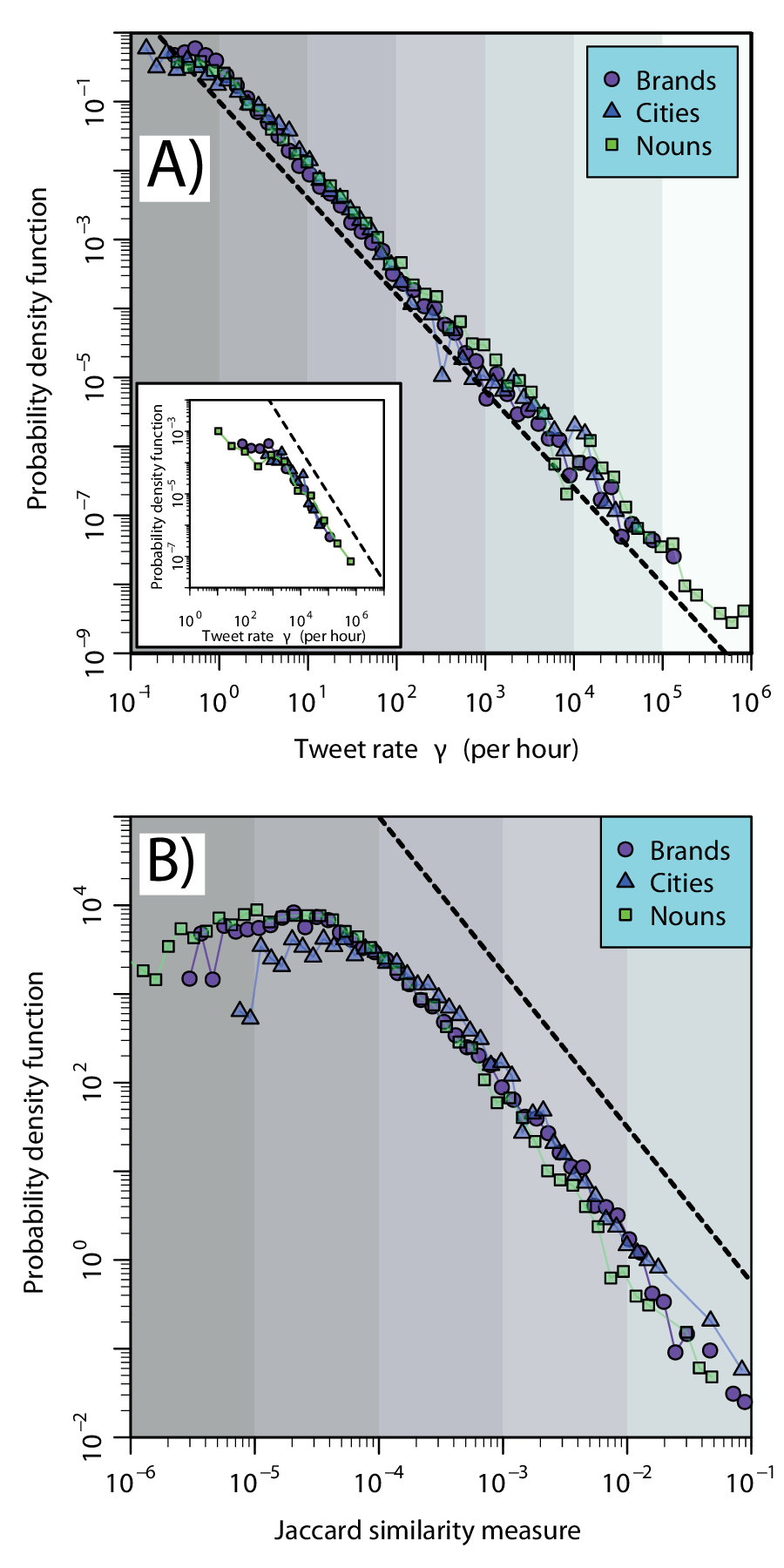}
\caption{
{Probability density function of tweet rates of pairs of international brands, major cities in the USA and common English nouns.} The distributions include rates of individual search terms. The violet circles correspond to brand names, the blue triangles to cities and the green squares to nouns. Note that the rates of the cities have been multiplied by 20 to allow for a direct comparison. The distributions of the rates are scale invariant over more than six orders of magnitude and have the same exponent $\alpha=1.40\pm 0.02$ (s.d.). The dashed line corresponds to $\alpha=1.4$. The inset shows distributions of tweet rates of single brands (purple circles), major US cities (blue triangles) and English nouns (green squares). For comparison we have inserted the same line as in the main panel and it is observed that the individual categories do not have the same scaling behavior.}\label{fig3}
\end{figure}



\begin{thebibliography}{10}%
\bibitem{B02} Albert R, Barabasi A.-L. (2002) Statistical mechanics of complex networks. {Rev. Mod. Phys.} { 74:} 47-97.
\bibitem{Borgatti2010} Borgatti SP,  Mehra A,  Brass DJ, Labianca G. (2009) Network analysis in the Social Sciences. {Science} {323:} 892-895.
\bibitem{Kitsak2010}
Kitsak M {\it et al.} (2010) Identification of influential spreaders in complex networks. {Nature Physics} {6:} 888-893.

\bibitem{RFFMV2010} Ratkiewskicz J, Fortunato S, Flammini F, Vespignani A (2010) Characterizing and modeling the dynamics of online popularity, Phys. Rev. Lett. 105: 158701
\bibitem{Twitter2011}
Mandavilli A (2011) Peer review: Trial by Twitter. {Nature} {469:}, 286-287.
\bibitem{Huberman2009}
Huberman BA,  Romero DM, Wu F (2009) Crowdsourcing, attention and productivity. {J. Inform. Sci.} {35:} 758-765.
\bibitem{KLPM10} Kwak H, Lee C, Park H, Moon S (2010) What is Twitter, a social network or a news media? Proceedings of the 19th international conference on World Wide Web, 591-600.

\bibitem{king2010} King G (2011) Ensuring the data-rich future of the social sciences. {Science} {331:} 719-721.
\bibitem{C10} Centola D (2010) The spread of behavior in an online social network experiment. {Science} {329:} 1194-1197.
\bibitem{CV07} R.L. Cilibrasi and P.M.B. Vitanyi (2007) The Google Similarity Distance, {IEEE T. Knowl. Data. En} 3, 370–383

\bibitem{rosvall} Rosvall M and Bergstrom C. T. (2008) Maps of random walks on complex networks reveal community structure. Proc. Natl. Acad. Sci. U.S.A. 105 1118-1123

\bibitem{Z49} Zipf G.K., Human behavior and the principle of least effort (Addison-Wesley, Cambridge, 1949)

\bibitem{ST05} Steyvers M, Tenenbaum JB (2005) The large-scale structure of semantic networks:
Statistical analyses and a model of semantic growth. Cognit Sci 29:41–78.

\bibitem{CB09} Cattuto C., Barrat A., Baldassarro A, Schehr G, and Loreto V. (2009) Collective dynamics of social annotation. Proc. Natl. Acad. Sci. U.S.A 26:10511-10515.


\bibitem{BRM10} Bernhardsson S, Correa da Rocha LE, Minnhagen P (2010) Size-dependent word frequencies and translational invariance of books, Physica A 389, 330-341.












\end{thebibliography}
\end{document}